\documentclass[12pt,superscriptaddress,showpacs,showkeys,twocolumn,prb,aps,reprint,a4paper,floatfix]{revtex4-1}

\usepackage{graphicx}
\usepackage{amsmath, amsthm, amssymb, amsfonts}
\usepackage{float}

\usepackage{natbib}
\usepackage{physics}
\usepackage[para,online,flushleft]{threeparttable} 
\usepackage{array}
\usepackage{ragged2e} 

\makeatletter
\newcommand*{\citenst}[2][]{%
  \begingroup
  \let\NAT@mbox=\mbox
  \let\@cite\NAT@citenum
  \let\NAT@space\NAT@spacechar
  \let\NAT@super@kern\relax
  \renewcommand\NAT@open{[}%
  \renewcommand\NAT@close{]}%
  \citet[#1]{#2}%
  \endgroup
}
\makeatother

\makeatletter
\newcommand*{\citenumns}[2][]{%
  \begingroup
  \let\NAT@mbox=\mbox
  \let\@cite\NAT@citenum
  \let\NAT@space\NAT@spacechar
  \let\NAT@super@kern\relax
  \renewcommand\NAT@open{[}
  \renewcommand\NAT@close{]}%
  \cite[#1]{#2}
  \endgroup
}
\makeatother
\usepackage[caption=false]{subfig}

\begin{document}
\title{Suppressed-gap millimetre wave kinetic inductance detectors using DC-bias current}
\author{Songyuan Zhao}
\email{sz311@cam.ac.uk}
\author{S. Withington}
\author{D. J. Goldie}
\author{C. N. Thomas}
\date{January 24, 2020}

\affiliation{Cavendish Laboratory, JJ Thomson Avenue, Cambridge CB3 OHE, United Kingdom.}

\begin{abstract}
\noindent In this study, we evaluate the suitability of using DC-biased aluminium resonators as low-frequency kinetic inductance detectors capable of operating in the frequency range of 50 - 120 GHz. Our analysis routine for supercurrent-biased resonators is based on the Usadel equations and gives outputs including density of states, complex conductivities, transmission line properties, and quasiparticle lifetimes. Results from our analysis confirm previous experimental observations on resonant frequency tuneability and retention of high quality factor. Crucially, our analysis suggests that DC-biased resonators demonstrate significantly suppressed superconducting density of states gap. Consequently these resonators have lower frequency detection threshold and are suitable materials for low-frequency kinetic inductance detectors.
\end{abstract}


\maketitle

\section{Introduction}

Kinetic inductance detectors (KIDs) are ultra-sensitive cryogenic detectors based on high-quality thin-film superconducting resonators. They can be straightforwardly multiplexed in the frequency domain, thereby allowing thousands of detectors to be read out by a common transmission line \citenumns{Day_2003, Jonas_review}. These detectors are readily fabricated using conventional ultra-high vacuum deposition techniques, and have demonstrated potential to be extensively applied in the areas of astronomy observations across the electromagnetic spectrum \citenumns{Monfardini_2010,Maloney_2010, Endo_2012, Mazin_2013}, neutrinoless double-beta decay experiments \citenumns{Battistelli2015, Cardani_2015}, dark matter search \citenumns{Golwala2008, Cornell_2018}, and general-purpose terahertz imaging \citenumns{Rowe_2016}.

The detection mechanism of KIDs requires the incoming photon to have sufficient energy to break a Cooper pair into quasiparticles, i.e. $\hbar \omega\geq \hbar \omega_{\mathrm{min}}=2\Delta_g$, where $\hbar$ is the reduced Planck constant, $\omega$ is the angular frequency of radiation, $\omega_{\mathrm{min}}$ is the angular frequency detection threshold, and $\Delta_g$ is the superconducting density of states energy gap. As a result of the frequency detection threshold, significant difficulty arises when the KIDs technology is applied to the detection of low-frequency millimetre wave signals, such as Cosmic Microwave Background radiation in the frequency range of $70 - 120\,\mathrm{GHz}$ \citenumns{Planck_hifi_2011, Catalano_2015}, low red-shifted  CO lines in the range of $100 - 110\,\mathrm{GHz}$ \citenumns{Cicone_CO_2012,Thomas_2014,Songyuan_2018}, and $\mathrm{O_2}$ rotation lines at $50-60\,\,{\mathrm {GHz}}$ for atmospheric profiling \citenumns{Mahfouf_2015,Aires_2015,Turner_2016}.  KIDs based on elemental superconductors are unable to simultaneously achieve high resonator quality factors as well as low frequency detection thresholds. Aluminium (Al), for example, has a strongly suppressed detector response below $100\,\mathrm{GHz}$ \citenumns{Catalano_2015}. To address the scientific need for low frequency KIDs, two alternative solutions have been explored: the usage of alloy superconductors, such as aluminium manganese (AlMn) \citenumns{Jones_2015} and titanium nitride (TiN) \citenumns{ Coiffard_2016_TiN, Giachero_Ti_TiN, Vissers_highQ_Ti_TiN_2013}, as well as the usage of multi-metallic-layer superconductors \citenumns{Catalano_2015, Songyuan_2018}. The alloy approach, in general, suffers from variations in material properties, even in a single deposition \citenumns{Vissers_2013}. Significant improvement in material uniformity has been demonstrated through the use of multilayer in conjunction with alloys \citenumns{Vissers_highQ_Ti_TiN_2013}. In contrast, the use of multi-elemental-metal-layer has the additional advantage of theoretical predictability in $\Delta_g$ through the application of the Usadel equations based on the BCS theory of superconductivity \citenumns{Usadel_1970, Songyuan_2018}, as well as predictability in electrical and optical properties through the application of the Mattis-Bardeen theory \citenumns{MattisBardeen_1958} (in contrast with TiN alloy which cannot be modelled using the Mattis-Bardeen theory \citenumns{Driessen_2012}). In this paper, we explore a third approach to the problem of low-frequency detection through the introduction of DC-bias currents to KIDs.

Various design schemes have been proposed and studied to introduce DC-bias currents to superconducting microwave resonators \citenumns{Chen_2011,Li_2013,Hao_2014,Bosman_2015,Vissers_2015,Adamyan_2016}. The context of these previous works include circuit quantum electro dynamics systems \citenumns{Wallraff_2004,Mallet_2009}, back-action-evading quantum measurement systems \citenumns{Hertzberg_2010}, and high sensitivity photo detection systems \citenumns{Day_2003, Jiansong_2012}. These studies are motived by the tuneable resonant frequencies \citenumns{Vissers_2015,Adamyan_2016} and the tuneable Josephson junction inductances \citenumns{Li_2013} of the biased devices. This present study is distinct from previous studies in exploiting the $\Delta_g$ suppression effect of bias currents. Introducing a DC-bias to a resonator such as KID without lowering its quality factor is an experimental challenge \citenumns{Li_2013,Adamyan_2016}. Hitherto, studies of DC-biased resonators have been mainly focused on their experimental realizations. Successful schemes have been developed for coplanar waveguides \citenumns{Chen_2011,Li_2013,Hao_2014,Bosman_2015,Vissers_2015} as well as for microstrip transmission lines \citenumns{Adamyan_2016}. In this work, we present a numerical analysis of DC-biased KIDs in terms of density of states, complex conductivity, transmission line quality factor, and quasiparticle lifetime. We explain various features in previous experimental studies such as frequency tuneability and high quality factors across a wide range of bias currents. Our results show that DC-biased KIDs have lower frequency detection thresholds due to the suppression of $\Delta_g$ in the presence of supercurrents. This opens up the possibility of using DC-biased KIDs to fulfil the current scientific need for low-frequency ultra-sensitive detector systems.

\section{Analysis Routine}
\begin{table}[ht]
\begin{threeparttable}
\caption{\label{tab:table1}Table of material properties.}
\begin{tabular}{b{0.40\linewidth} b{0.26\linewidth}}
\toprule
 & \textrm{Aluminium} \\
\colrule
$T_{\mathrm {c}}$ ($\mathrm{K}$) & 1.20\tnote{a} \\
$\sigma_{\mathrm{N}}$ ($\mathrm{MS/m}$) \tnote{b} & 132\tnote{a}\\
$N_0$ ($10^{47}$/$\textrm{J}\,\textrm{m}^3$) & 1.45\tnote{c}\\
$D$ ($\mathrm{m^2s^{-1}}$) & 35\tnote{d} \\
$\xi$ ($\mathrm{nm}$) & 189\tnote{e}\\
$\Theta_D$ ($\mathrm{K}$) & 423\tnote{f}\\
$\tau_0$ ($\mathrm{ns}$) & 395\tnote{g}\\
\toprule
\end{tabular}
\begin{tablenotes}[flushleft]
\RaggedRight
\footnotesize
\item[a] $T_{\mathrm {c}}$ is the superconductor critical temperature. Value is measured. \\
\item[b] $\sigma_{\mathrm{N}}$ is the normal state conductivity. \\
\item[c] $N_0$ is the normal state electron density of states, and is calculated from the free electron model \citenumns{Ashcroft_1976}. \\
\item[d] Diffusivity constant $D$ is calculated using $D = \sigma_{N}/(N_{0}e^2)$ \citenumns{Martinis_2000}. \\
\item[e] Coherence length $\xi$ is calculated using $\xi=[{\hbar D/(2\pi k_\mathrm {B} T_\mathrm {c}})]^{1/2}$ \citenumns{Brammertz2004}, where $k_\mathrm {B}$ is the Boltzmann constant. \\
\item[f] $\Theta_D$ is the Debye temperature, and is given by $k_B\Theta_D=\hbar \omega_D$, where $\omega_D$ is the Debye frequency. Value is taken from \citenumns{Gladstone}. \\
\item[g] $\tau_0$ is the characteristic electron-phonon coupling time. Value is taken from \citenumns{Parlato2005}.
\end{tablenotes}
\end{threeparttable}
\end{table}

In order to establish if the frequency detection threshold of a KID can be lowered through the introduction of a DC supercurrent, and if the resulting biased resonator retains desirable properties to operate as a KID, we have developed an analysis routine based on the Usadel equations. The Usadel equations \citenumns{Usadel_1970} are a pair of diffusive limit differential equations based on the microscopic BCS theory of superconductivity. The equations have been extensively applied to the analysis of thin-film superconducting devices \citenumns{Martinis_2000,Brammertz2004,Buzdin_2005} and the theoretical predictions have demonstrated excellent agreement with experimental results \citenumns{Radovic_1991,Anthore_2003,Songyuan_2018_Tc}. Our analysis routine consists of the following components:
\begin{enumerate}
  \item The Usadel equations are solved self-consistently to obtain the superconducting Green's functions and superconducting densities of states (DoSs) in the presence of supercurrent.
  \item Nam's equations \citenumns{Nam_1967} are integrated to compute the complex conductivities $\sigma = \sigma_1-i\sigma_2$ from the superconducting DoSs.
  \item Complex surface impedances $Z_s = R_s + j\omega L_s$ are computed using
\begin{align}
Z_s=\left(\frac{j\omega\mu_0}{\sigma}\right)^{1/2}\operatorname{coth}[(j\omega\mu_0\sigma)^{1/2}t] , \label{eq:Zs}
\end{align}
  where $j$ is the unit imaginary number, $t$ is the thickness of the superconducting film, $\mu_0$ is the vacuum permeability, and $\omega$ is the angular frequency of the signal of interest \citenumns{Kerr1996,Withington_1995}.
  \item Transmission line properties are computed using suitable conformal mapping results for the specific resonator geometry \citenumns{Songyuan_transmission_lines_2018}. The series impedance and shunt admittance are given by
  \begin{align}
\label{eqn:Z}
Z &=j(k_0\eta_0)g_1 +2\sum_{n}g_{2,n}Z_{s,n} = R+j\omega L\\
\label{eqn:Y}
Y &=j\left(\frac{k_0}{\eta_0}\right)\left(\frac{\epsilon_{fm}}{g_1}\right)=G+j\omega C \, ,
\end{align}
where $k_0$ is the free-space wavenumber, $\eta_0$ is the impedance of free-space, subscript $n$ identifies superconductor surfaces, which are upper, lower conductor surfaces, and ground surfaces of the transmission line, denoted by subscripts $u,l$, and $g$ respectively, $\epsilon_{fm}$ is the effective modal dielectric constant, which is given by existing normal conductor transmission line theories, for example \citenumns{Edwards_1976, Gupta_1996}. $g_1$ and $g_2$ are geometric factors which are calculated using appropriate conformal mapping results from \citenumns{Songyuan_transmission_lines_2018}. $R$, $G$, $L$, $C$ are the resistance, conductance, inductance, and capacitance, per unit length, respectively.
  \item Quasiparticle recombination lifetimes are computed using the low-energy expression given in \citenumns{Golubov1994}, and the energy-averaged recombination lifetimes are then calculated according to the weighted-average procedure given in \citenumns{Songyuan_2018}.
\end{enumerate}
Explicit equations used in each numerical component are given with more details in \citenumns{songyuan2019_nonlinear, Songyuan_2018}. The results in the next section are obtained by applying this analysis routine to DC-biased coplanar waveguide (CPW) KIDs based on Al. Here we have adopted a basic model for KIDs which assumes that the same superconducting material (Al) is responsible both for photon absorption as well as for readout resonance. The properties of Al used in this analysis are taken from a previous study \citenumns{Songyuan_2018_Tc}, and are shown in Table~\ref{tab:table1}. The modelled CPW geometry has inner half-width $a=1.0\,\mathrm{\mu m}$, gap width $b-a=0.5\,\mathrm{\mu m}$, thickness $t=20\,\mathrm{nm}$, dielectric height $h=225\,\mathrm{\mu m}$, dielectric constant $\epsilon_r=11.7$, and dielectric quality factor $Q_{\epsilon}=10^5$ in accordance with measured values for silicon at cryogenic temperatures \citenumns{Connell_2008}. KIDs are typically read out by a microwave probe operating at $1-10\,\mathrm{GHz}$ \citenumns{Day_2003,Jiansong_2012}. As such, results in the next section are calculated using a readout frequency $f_r$ of $10\,\mathrm{GHz}$.

\section{Results}

\begin{figure}[ht]
\includegraphics[width=8.6cm]{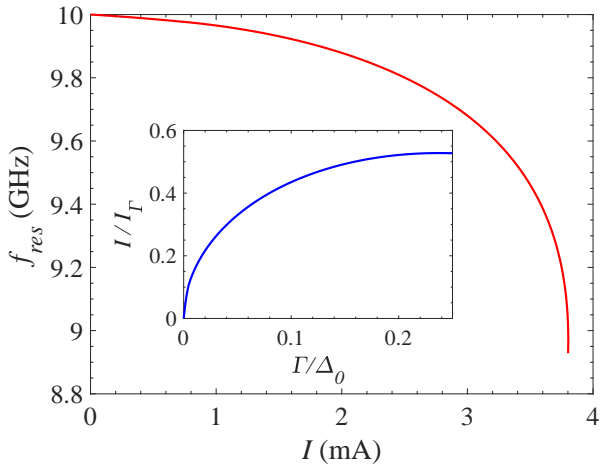}
\caption{\label{fig:res_Freq} Al CPW resonant frequency $f_{\mathrm{res}}$ against supercurrent $I$. Inset: scaled supercurrent $I/I_{\Gamma}$ against supercurrent depairing factor $\Gamma/\Delta_0$.}
\end{figure}

Figure \ref{fig:res_Freq} shows resonant frequency $f_{\mathrm{res}}$ against supercurrent $I$ for the Al CPW with dimensions described in the previous section. This tuneability in resonant frequency is the subject and motivation of previous studies on DC-biased resonators \citenumns{Vissers_2015,Adamyan_2016}. It is important to note that the quantitative dependence is specific to the geometry of the resonator. The DC-bias affects only the kinetic inductance but not the geometric inductance. As such, a device with a higher kinetic inductance to geometric inductance ratio will, in general, demonstrate greater maximum tuneability. Design wise, transmission line theories such as \citenumns{Songyuan_transmission_lines_2018} can be used to improve this ratio.

In this study we express bias currents in terms of normalized supercurrent depairing factors $\Gamma/\Delta_0$. This is because $\Gamma/\Delta_0$ comes out naturally from the Usadel equations and is not device geometry/material dependent. The conversion between  $\Gamma/\Delta_0$ and physical supercurrent $I$ can be done using equation (7) of \citenumns{Anthore_2003}. We have plotted this conversion in the inset of figure \ref{fig:res_Freq}, which shows scaled supercurrent $I/I_{\Gamma}$ against normalized supercurrent depairing factor $\Gamma/\Delta_0$. The current scaling factor is given by $I_{\Gamma}=\sqrt{2}S\Delta_0\sigma_N/(e\xi)$, where $S$ is the cross-section area of the resonator, $\sigma_N$ is the normal state conductivity, $e$ is the electron charge, and $\xi$ is the material coherence length. The critical current is given by $I_c\approx 0.53 \, I_{\Gamma}$ \citenumns{Anthore_2003}. For $T\ll T_c$, $I=0\,\mathrm{A}$ when $\Gamma/\Delta_0=0$ and $I=I_c$ when $\Gamma/\Delta_0=0.25$. For Al, using parameters given in Table~\ref{tab:table1}, $I_{\Gamma}/S\approx 1.8\times 10^{11} \mathrm{A/m^2}$ and $I_{c}/S\approx 9.6\times 10^{10} \mathrm{A/m^2}$. For an Al CPW with geometry described in the previous section, $I_{\Gamma}=7.2\,\mathrm{mA}$ and $I_{c}= 3.8\,\mathrm{mA}$. Three particular values of $\Gamma/\Delta_0$ are used consistently across different figures: $\Gamma/\Delta_0 = 5.0\times10^{-3}$ demonstrates device behaviour at very low bias current, $\Gamma/\Delta_0 = 2.0\times10^{-1}$ demonstrates device behaviour at close to critical current, and $\Gamma/\Delta_0 = 1.0\times10^{-1}$ demonstrates device behaviour at intermediate current values.

\begin{figure}[ht]
\includegraphics[width=8.6cm]{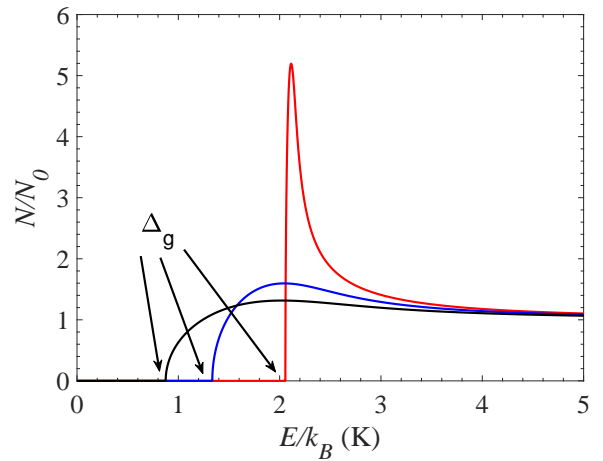}
\caption{\label{fig:DoSes} Al superconducting density of states $N/N_0$ against energy $E/k_{B}$ at temperature $T=0.01\,\mathrm{K}$ for different values of supercurrent depairing factor $\Gamma/\Delta_0$. Red line: $\Gamma/\Delta_0 = 5.0\times10^{-3}$; blue line: $\Gamma/\Delta_0 = 1.0\times10^{-1}$; black line: $\Gamma/\Delta_0 = 2.0\times10^{-1}$. Density of states gaps are labelled $\Delta_g$. }
\end{figure}
Figure \ref{fig:DoSes} shows the superconducting density of states $N/N_0$ of Al against energy $E/k_{B}$ for different values of normalized supercurrent depairing factor $\Gamma/\Delta_0$, where $\Delta_0=1.764\,k_B T_c$. As seen in the figure, the shape of the DoS is broadened and the DoS gap is suppressed in the presence of bias current. This effect on the superconducting DoSs has been observed in previous experimental data \citenumns{Anthore_2003}.

\begin{figure}[ht]
\includegraphics[width=8.6cm]{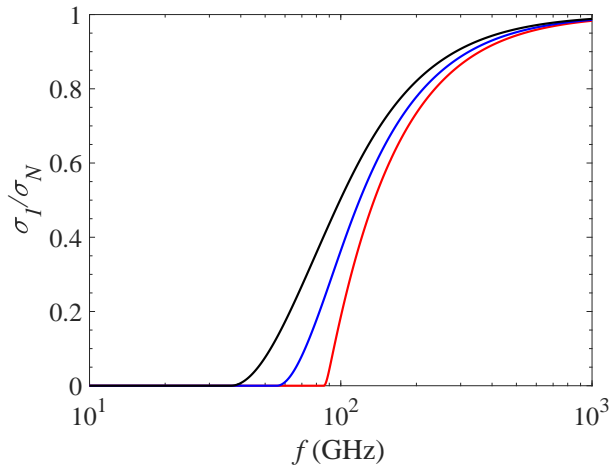}
\caption{\label{fig:con_Real} Al dissipative conductivity $\sigma_1/\sigma_N$ against frequency $f$ at temperature $T=0.01\,\mathrm{K}$ for different values of supercurrent depairing factor $\Gamma/\Delta_0$. Red line: $\Gamma/\Delta_0 = 5.0\times10^{-3}$; blue line: $\Gamma/\Delta_0 = 1.0\times10^{-1}$; black line: $\Gamma/\Delta_0 = 2.0\times10^{-1}$.}
\end{figure}

\begin{figure}[ht]
\includegraphics[width=8.6cm]{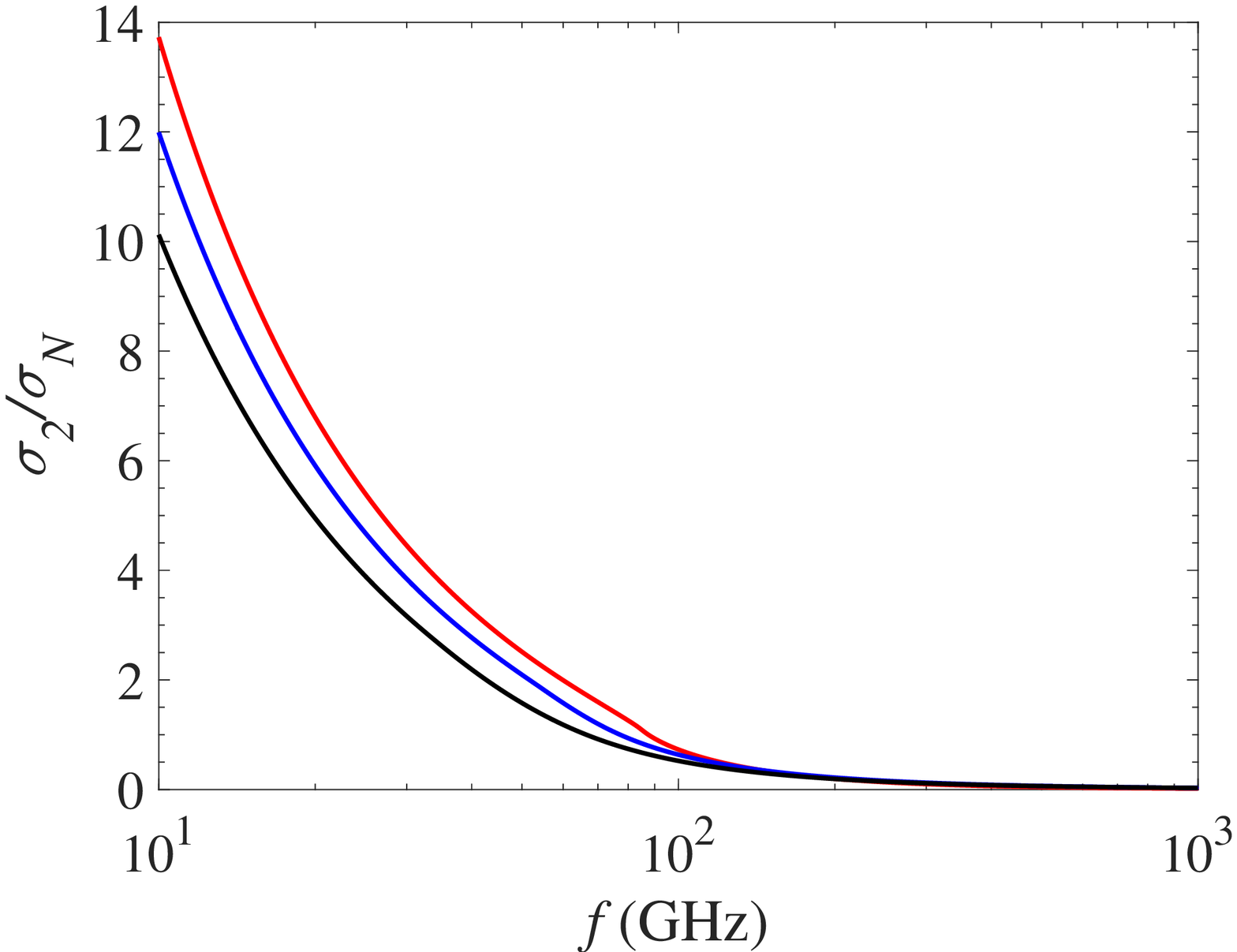}
\caption{\label{fig:con_Imag} Al reactive conductivity $\sigma_2/\sigma_N$ against frequency $f$ at temperature $T=0.01\,\mathrm{K}$ for different values of supercurrent depairing factor $\Gamma/\Delta_0$. Red line: $\Gamma/\Delta_0 = 5.0\times10^{-3}$; blue line: $\Gamma/\Delta_0 = 1.0\times10^{-1}$; black line: $\Gamma/\Delta_0 = 2.0\times10^{-1}$.}
\end{figure}

Figure \ref{fig:con_Real} and figure \ref{fig:con_Imag} show the real (dissipative) and imaginary (reactive) components of the complex conductivity respectively against frequency, for different values of DC-bias. The shift in reactive conductivity $\sigma_2/\sigma_N$ at readout frequencies is responsible for the shift in resonant frequency. As seen in figure \ref{fig:con_Imag}, $\sigma_2/\sigma_N$ is suppressed in the presence of supercurrent. This in turn results in a boost in surface inductance $L_s$ and transmission line inductance $L$ through equation (\ref{eq:Zs},\ref{eqn:Z}). The increased $L$ then results in a lowering of the resonant frequency, central to the operation of frequency tuneable resonators experimentally demonstrated in \citenumns{Vissers_2015,Adamyan_2016}. The shift in the gap of dissipative conductivity $\sigma_1/\sigma_N$, on the other hand, is responsible for the lowering of the frequency detection threshold as $\sigma_1/\sigma_N$ is the absorption ratio for electromagnetic radiation \citenumns{Tinkham_1994}. As seen in figure \ref{fig:con_Real}, photon detection is possible at lower frequencies in the presence of supercurrent.

\begin{figure}[ht]
\includegraphics[width=8.6cm]{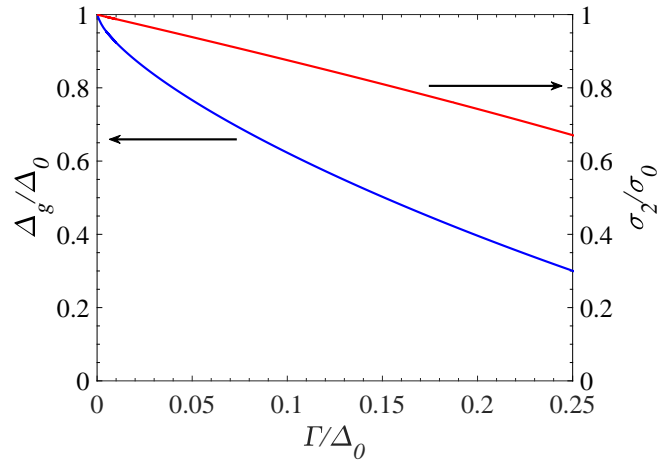}
\caption{\label{fig:Gap_Suppression} Red line: Al normalized reactive conductivity $\sigma_2/\sigma_0$ against supercurrent depairing factor $\Gamma/\Delta_0$ at readout frequency $f_{r}=10\,\mathrm{GHz}$; blue line: Al normalized density of states gap $\Delta_g/\Delta_0$ against supercurrent depairing factor $\Gamma/\Delta_0$. }
\end{figure}

Figure \ref{fig:Gap_Suppression} shows the dependence of normalized $\sigma_2/\sigma_0$ (red line) and normalized $\Delta_g/\Delta_0$ (blue line) against $\Gamma/\Delta_0$ at readout frequency $f_{r}=10\,\mathrm{GHz}$, with the normalization factor defined as $\sigma_0=\sigma_2(\Gamma=0\,\mathrm{K})\approx\pi\Delta_0/(\hbar\omega)$ and $\Delta_0=\Delta_g(\Gamma=0\,\mathrm{K})z\approx1.764\,k_B T_c$. As seen in the figure, the extent of shift in $\Delta_g/\Delta_0$ is much greater compared to $\sigma_2/\sigma_0$. This is because the DoS gaps are the furthest shifted points on the DoSs, whereas the conductivities are energy integrals over functions of DoSs. From figure \ref{fig:Gap_Suppression}, we can predict the amount of gap suppression given a known tuneability on the resonant frequency. For example, \citenumns{Vissers_2015} reports a $4\%$ tuneability in frequency. Assuming the applicability of the Mattis-Bardeen theory, and that the kinetic inductance dominates the contributions to total device inductance, we estimate $\sigma_2/\sigma_0\approx 0.92$. Using figure \ref{fig:Gap_Suppression}, we expect $\Delta_g/\Delta_0\approx0.72$, i.e. the frequency detection threshold is expected to be suppressed by almost $25\%$. If similar or better performance could be translated to Al resonators, we expect DC-biasing to greatly extend the application of Al KIDs.

\begin{figure}[ht]
\includegraphics[width=8.6cm]{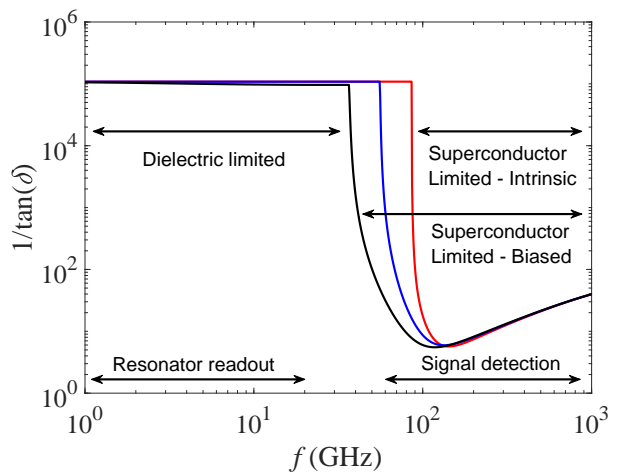}
\caption{\label{fig:Trans_Line_Quality} Al CPW quality factor $1/\mathrm{tan} \delta$ against frequency $f$ at temperature $T=0.01\,\mathrm{K}$ for different values of supercurrent depairing factor $\Gamma/\Delta_0$. Red line: $\Gamma/\Delta_0 = 5.0\times10^{-3}$; blue line: $\Gamma/\Delta_0 = 1.0\times10^{-1}$; black line: $\Gamma/\Delta_0 = 2.0\times10^{-1}$.}
\end{figure}

One important consideration in the design of low frequency KIDs for applications requiring high detector sensitivity is the overall device quality factor \citenumns{Jonas_review}. Figure \ref{fig:Trans_Line_Quality} shows the transmission line quality factor against frequency for different bias supercurrents at $T=0.01\,\mathrm{K}$. As seen in the figure, the presence of supercurrent minimally affects the overall quality factor at readout frequencies. This is because, at low temperatures, the superconductor quality factor from the Mattis-Bardeen theory far exceeds that of the CPW dielectric. As a result, the transmission line quality factor is dielectric limited \citenumns{Adamyan_2016}. This insensitivity of the transmission line quality factor to bias current has been observed in \citenumns{Vissers_2015}, and is important in allowing DC-biased resonators to be used as high sensitivity detectors. The frequency dependence of the quality factor can be broadly divided into a low-frequency region which is limited by the dielectric quality factor, and a high-frequency region which is limited by the superconductor quality factor above the DoS gap. The low-frequency region is relevant to KIDs readout, whereas the high-frequency region is relevant to photon detection.

\begin{figure}[ht]
\includegraphics[width=8.6cm]{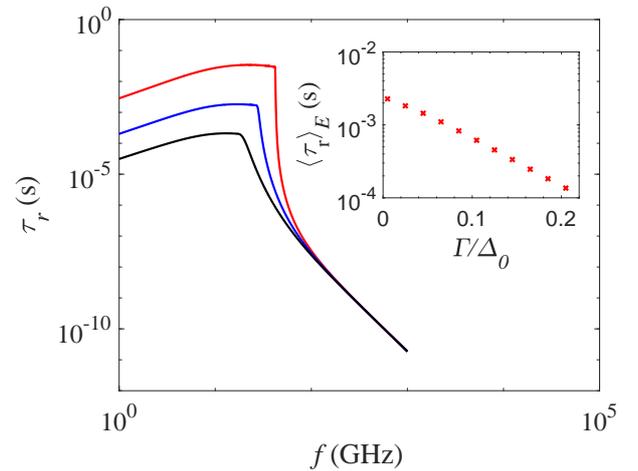}
\caption{\label{fig:Lifetime} Al quasiparticle lifetime $\tau_r$ against frequency $f$ at temperature $T=0.01\,\mathrm{K}$ for different values of supercurrent depairing factor $\Gamma/\Delta_0$. Red line: $\Gamma/\Delta_0 = 5.0\times10^{-3}$; blue line: $\Gamma/\Delta_0 = 1.0\times10^{-1}$; black line: $\Gamma/\Delta_0 = 2.0\times10^{-1}$. Inset: energy averaged quasiparticle lifetime $\langle\tau_{\mathrm{r}}\rangle_{E}$ against supercurrent depairing factor $\Gamma/\Delta_0$. }
\end{figure}

Another important consideration in evaluating the suitability of DC-biased KIDs is the quasiparticle recombination lifetime which governs the trade-off between detector response time and recombination noise \citenumns{Leduc_2010, Jonas_review}. Figure \ref{fig:Lifetime} shows the recombination lifetime $\tau_r$ against frequency for different values of bias current. The inset shows the energy averaged recombination lifetime $\langle\tau_{\mathrm{r}}\rangle_{E}$ against $\Gamma/\Delta_0$. This calculation is performed at $T=0.15\,\mathrm{K}$, close to the saturation point of quasiparticle lifetime for Al \citenumns{Barends_Lifetime_2008}. The presence of supercurrent decreases the recombination lifetime across the energy spectrum. The inset shows that the energy-averaged lifetime has an inverse exponential dependence on the depairing factor, which is proportional to the squared current. At $T<<T_c$, the recombination lifetime also has an inverse exponential dependence on $T$ \citenumns{Kaplan_1976}. This suggests the possibility of interpreting the effect of depairing current on quasiparticle lifetime as raising the effective temperature. It is important to note that the lifetime calculation presented here assumes BCS-like behaviour from the superconductors. It is well documented that quasiparticle lifetime, as well as quasiparticle number, plateaus and deviates from BCS predictions at low temperatures $T/T_c<0.15$ \citenumns{Barends_Lifetime_2008,deVisser2012}. Future experimental studies should be conducted to determine the low temperature lifetime behaviour of DC-biased KIDs: whether the DC-bias lowers the overall low-temperature lifetime curve (thereby lowering the low temperature saturation plateau), or whether the DC-bias effect can be account by an adjusted effective temperature along an existing lifetime curve (without changing the low temperature plateau height). In the first case more optimization may be needed to suit different applications, and in the second case the device can be operated simply as regular Al KIDs in terms of recombination lifetime.

\begin{figure}[ht]
\includegraphics[width=8.6cm]{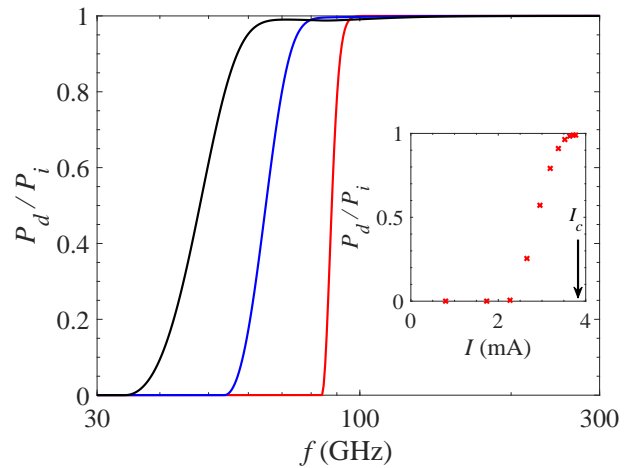}
\caption{\label{fig:Power_dissipated} Fractional power dissipated $P_{d}/P_{i}$ by a DC-biased Al CPW resonator with length $l=5\,\mathrm{mm}$ fed by a non-biased Al CPW of the same dimensions against frequency of signal $f$. Red line: $\Gamma/\Delta_0 = 5.0\times10^{-3}$; blue line: $\Gamma/\Delta_0 = 1.0\times10^{-1}$; black line: $\Gamma/\Delta_0 = 2.0\times10^{-1}$. Inset: fractional power dissipated $P_{d}/P_{i}$ against magnitude of bias current $I$ at frequency $f=70\,\mathrm{GHz}$.}
\end{figure}

To illustrate the gap-suppression characteristic of a biased KID, we have calculated the fractional power dissipated $P_{d}/P_{i}$ by a DC-biased Al CPW with length $l=5\,\mathrm{mm}$ fed by a non-biased Al CPW of the same dimensions, terminated at a matched load. Here $P_{i}$ is the incident power and $P_{d}$ is the dissipated power. $P_{d}/P_{i}$ is calculated from applying the transmission line dissipation propagation constant $\alpha$ over the length of the resonator \citenumns{Songyuan_transmission_lines_2018}, after taking into account the reflection off the interface between the non-biased feedline and the biased resonator. Figure \ref{fig:Power_dissipated} shows the signal frequency $f$ dependence of $P_{d}/P_{i}$. As seen in the figure, the power dissipated in the resonator increases sharply above the DoS gaps. The dissipated power $P_{d}$ increases with transmission line length $l$ according to $P_{d}/P_{t} = 1-e^{-2\alpha l}$ \citenumns{pozar2011microwave,Songyuan_transmission_lines_2018}, where $P_{t}$ is the transmitted power across the boundary between the non-biased CPW and the biased CPW. This scaling relation has the important consequence that the steepness of the rise in dissipation with frequency can be further increased through the use of longer resonator lengths. The inset of figure \ref{fig:Power_dissipated} shows the bias current $I$ dependence of $P_{d}/P_{i}$ at frequency $f=70\,\mathrm{GHz}$. The power dissipation is initially nearly zero due to the absence of significant superconductor loss at sub-pair-breaking frequency. As $I$ increases, pair-breaking frequency is reduced, and significant dissipation sets in when the pair-breaking frequency is suppressed below the signal frequency.


\section{Conclusion}
We have studied theoretically and numerically the feasibility of operating KIDs below their unbiased density of states gaps through DC-biasing the superconducting thin-films used for photon detection. Our numerical analysis is based on the Usadel equations and evaluates detector performance in terms of density of states, complex conductivities, transmission line quality factors, and quasiparticle lifetimes. Our results confirm previous experimental observations on the tuneability of the resonant frequencies and the high quality factors of DC-biased resonators. Our analysis further predicts significant suppression in the frequency threshold of photon detection in the presence of DC-bias current. This phenomenon allows DC-biased resonators to be used as KIDs to fulfil the scientific need for high sensitivity, low-frequency threshold photon detectors. One important effect observed by previous experimental studies is the sharp onset of dissipation at high current values (but below the theoretical critical current), resulting in deviations from ideal calculations \citenumns{songyuan2019_nonlinear}. This phenomenon places a limit on the maximum tuneability of resonant frequency and detection threshold \citenumns{Vissers_2015,Adamyan_2016}. Future investigations should be conducted to extend the range of resonant frequency and detection threshold tuneability before the onset of significant dissipation, for examples, by decreasing the cross-sectional dimensions of the resonators \citenumns{songyuan2019_nonlinear}. Lastly, our numerical analysis shows that low-frequency $50 - 120\,\mathrm{GHz}$ Al KIDs with high quality factors can be made by incorporating DC-bias schemes. In view of this, experimental realizations of DC-biased Al KIDs should be conducted to directly measure the suppression of frequency detection thresholds and to characterise their detector performance.

\bibliographystyle{h-physrev}
\bibliography{library}

\begin{thebibliography}{10}

\bibitem{Day_2003}
P.~K. Day, H.~G. LeDuc, B.~A. Mazin, A.~Vayonakis, and J.~Zmuidzinas,
\newblock Nature {\bf 425}, 817 (2003).

\bibitem{Jonas_review}
J.~Zmuidzinas,
\newblock {Annu. Rev. Condens. Matter Phys.} {\bf {3}}, 169 ({2012}).

\bibitem{Monfardini_2010}
{Monfardini, A.} {\em et~al.},
\newblock A\&A {\bf 521}, A29 (2010).

\bibitem{Maloney_2010}
P.~R. Maloney {\em et~al.},
\newblock {MUSIC for sub/millimeter astrophysics},
\newblock in {\em Millimeter, Submillimeter, and Far-Infrared Detectors and
  Instrumentation for Astronomy V}, edited by W.~S. Holland and J.~Zmuidzinas
  Vol. 7741, pp. 124 -- 134, International Society for Optics and Photonics,
  SPIE, 2010.

\bibitem{Endo_2012}
A.~Endo {\em et~al.},
\newblock Journal of Low Temperature Physics {\bf 167}, 341 (2012).

\bibitem{Mazin_2013}
B.~A. Mazin {\em et~al.},
\newblock Publications of the Astronomical Society of the Pacific {\bf 125},
  1348 (2013).

\bibitem{Battistelli2015}
E.~S. Battistelli {\em et~al.},
\newblock The European Physical Journal C {\bf 75}, 353 (2015).

\bibitem{Cardani_2015}
L.~Cardani {\em et~al.},
\newblock Applied Physics Letters {\bf 107}, 093508 (2015),
  https://doi.org/10.1063/1.4929977.

\bibitem{Golwala2008}
S.~Golwala {\em et~al.},
\newblock Journal of Low Temperature Physics {\bf 151}, 550 (2008).

\bibitem{Cornell_2018}
B.~D. Cornell,
\newblock {\em A Dark Matter Search Using the Final SuperCDMS Soudan Dataset
  and the Development of a Large-Format, Highly-Multiplexed,
  Athermal-Phonon-Mediated Particle Detector},
\newblock PhD thesis, California Institute of Technology, 2018.

\bibitem{Rowe_2016}
S.~Rowe {\em et~al.},
\newblock Review of Scientific Instruments {\bf 87}, 033105 (2016),
  https://aip.scitation.org/doi/pdf/10.1063/1.4941661.

\bibitem{Planck_hifi_2011}
{Planck HiFi Core Team},
\newblock {Astron. Astrophys.} {\bf {536}}, {A4} ({2011}).

\bibitem{Catalano_2015}
A.~Catalano {\em et~al.},
\newblock {Astron. Astrophys.} {\bf {580}}, {A15} ({2015}).

\bibitem{Cicone_CO_2012}
C.~Cicone {\em et~al.},
\newblock {Astron. Astrophys.} {\bf {543}}, {A99} ({2012}).

\bibitem{Thomas_2014}
C.~N. Thomas {\em et~al.},
\newblock ({2014}), {astro-ph.IM/1401.4395v1}.

\bibitem{Songyuan_2018}
S.~Zhao, D.~J. Goldie, S.~Withington, and C.~N. Thomas,
\newblock Supercond. Sci. Tech. {\bf 31}, 015007 (2018).

\bibitem{Mahfouf_2015}
J.~Mahfouf {\em et~al.},
\newblock {Q. J. R. Meteorol. Soc.} {\bf {141}}, 3268 ({2015}).

\bibitem{Aires_2015}
F.~Aires {\em et~al.},
\newblock {J. Geophys. Res. Atmos.} {\bf {120}}, 11334 ({2015}).

\bibitem{Turner_2016}
E.~C. Turner {\em et~al.},
\newblock {Atmos. Meas. Tech.} {\bf {9}}, 5461 ({2016}).

\bibitem{Jones_2015}
G.~Jones {\em et~al.},
\newblock astro-ph.IM , 1701.08461v2 (2017).

\bibitem{Coiffard_2016_TiN}
G.~Coiffard {\em et~al.},
\newblock {J. Low Temp. Phys.} {\bf {184}}, 654 ({2016}).

\bibitem{Giachero_Ti_TiN}
A.~Giachero {\em et~al.},
\newblock J. Low Temp. Phys. {\bf 176}, 155 (2014).

\bibitem{Vissers_highQ_Ti_TiN_2013}
M.~R. Vissers {\em et~al.},
\newblock Appl. Phys. Lett. {\bf 102} ({2013}).

\bibitem{Vissers_2013}
M.~R. Vissers {\em et~al.},
\newblock Thin Solid Films {\bf 548}, 485  (2013).

\bibitem{Usadel_1970}
K.~D. Usadel,
\newblock Phys. Rev. Lett. {\bf 25}, 507 (1970).

\bibitem{MattisBardeen_1958}
D.~Mattis and J.~Bardeen,
\newblock {Phys. Rev.} {\bf {111}}, 412 ({1958}).

\bibitem{Driessen_2012}
E.~F.~C. Driessen, P.~C. J.~J. Coumou, R.~R. Tromp, P.~J. de~Visser, and T.~M.
  Klapwijk,
\newblock Phys. Rev. Lett. {\bf 109}, 107003 (2012).

\bibitem{Chen_2011}
F.~Chen, A.~J. Sirois, R.~W. Simmonds, and A.~J. Rimberg,
\newblock Applied Physics Letters {\bf 98}, 132509 (2011),
  https://doi.org/10.1063/1.3573824.

\bibitem{Li_2013}
S.-X. Li and J.~B. Kycia,
\newblock Applied Physics Letters {\bf 102}, 242601 (2013),
  https://doi.org/10.1063/1.4808364.

\bibitem{Hao_2014}
Y.~Hao, F.~Rouxinol, and M.~D. LaHaye,
\newblock Applied Physics Letters {\bf 105}, 222603 (2014),
  https://doi.org/10.1063/1.4903777.

\bibitem{Bosman_2015}
S.~J. Bosman, V.~Singh, A.~Bruno, and G.~A. Steele,
\newblock Applied Physics Letters {\bf 107}, 192602 (2015),
  https://doi.org/10.1063/1.4935346.

\bibitem{Vissers_2015}
M.~R. Vissers {\em et~al.},
\newblock Applied Physics Letters {\bf 107}, 062601 (2015),
  https://doi.org/10.1063/1.4927444.

\bibitem{Adamyan_2016}
A.~A. Adamyan, S.~E. Kubatkin, and A.~V. Danilov,
\newblock Applied Physics Letters {\bf 108}, 172601 (2016),
  https://doi.org/10.1063/1.4947579.

\bibitem{Wallraff_2004}
A.~Wallraff {\em et~al.},
\newblock Nature {\bf 431}, 162 (2004).

\bibitem{Mallet_2009}
F.~Mallet {\em et~al.},
\newblock Nature Physics {\bf 5}, 791 (2009).

\bibitem{Hertzberg_2010}
J.~B. Hertzberg {\em et~al.},
\newblock Nature Physics {\bf 6}, 213 (2010).

\bibitem{Jiansong_2012}
J.~Gao,
\newblock {\em The physics of superconducting microwave resonators},
\newblock PhD thesis, California Institute of Technology, 2008.

\bibitem{Ashcroft_1976}
N.~Ashcroft and N.~Mermin,
\newblock {\em {Solid State Physics}} (Saunders College, Philadelphia, 1976).

\bibitem{Martinis_2000}
J.~M. Martinis, G.~Hilton, K.~Irwin, and D.~Wollman,
\newblock Nucl. Instrum. Meth. A {\bf 444}, 23  (2000).

\bibitem{Brammertz2004}
G.~Brammertz {\em et~al.},
\newblock J. Appl. Phys. {\bf 90}, 355 (2001).

\bibitem{Gladstone}
G.~Gladstone, M.~A. Jensen, and J.~R. Schrieffer,
\newblock Superconductivity in the transition metals: theory and measurement,
\newblock in {\em Superconductivity}, edited by R.~D. Parks, pp. 665--816,
  Marcel Dekker, New York, 1969.

\bibitem{Parlato2005}
L.~Parlato {\em et~al.},
\newblock {Supercond. Sci. Technol.} {\bf {18}}, 1244 ({2005}).

\bibitem{Buzdin_2005}
A.~I. Buzdin,
\newblock Rev. Mod. Phys. {\bf 77}, 935 (2005).

\bibitem{Radovic_1991}
Z.~Radovi\ifmmode~\acute{c}\else \'{c}\fi{}, M.~Ledvij,
  L.~Dobrosavljevi\ifmmode \acute{c}\else
  \'{c}\fi{}-Gruji\ifmmode~\acute{c}\else \'{c}\fi{}, A.~I. Buzdin, and J.~R.
  Clem,
\newblock Phys. Rev. B {\bf 44}, 759 (1991).

\bibitem{Anthore_2003}
A.~Anthore, H.~Pothier, and D.~Esteve,
\newblock Physical review letters {\bf 90}, 127001 (2003).

\bibitem{Songyuan_2018_Tc}
S.~Zhao, D.~J. Goldie, C.~N. Thomas, and S.~Withington,
\newblock Superconductor Science and Technology {\bf 31}, 105004 (2018).

\bibitem{Nam_1967}
S.~B. Nam,
\newblock {Phys. Rev.}  ({1967}).

\bibitem{Kerr1996}
A.~R. Kerr,
\newblock National Radio Astronomy Observatory Report No. 245, 1996
  (unpublished).

\bibitem{Withington_1995}
G.~Yassin and S.~Withington,
\newblock J. Phys. D Appl. Phys. {\bf 28}, 1983 (1995).

\bibitem{Songyuan_transmission_lines_2018}
S.~Zhao, S.~Withington, D.~J. Goldie, and C.~N. Thomas,
\newblock Superconductor Science and Technology {\bf 31}, 085012 (2018).

\bibitem{Edwards_1976}
T.~C. Edwards and R.~P. Owens,
\newblock IEEE T. Microw. Theory Techn. {\bf 24}, 506 (1976).

\bibitem{Gupta_1996}
K.~Gupta,
\newblock {\em Microstrip Lines and Slotlines} (Artech House, 1996).

\bibitem{Golubov1994}
A.~A. Golubov {\em et~al.},
\newblock {Phys. Rev. B} {\bf {49}}, 12953 ({1994}).

\bibitem{songyuan2019_nonlinear}
S.~Zhao, S.~Withington, D.~J. Goldie, and C.~N. Thomas,
\newblock Journal of Low Temperature Physics  (2020).

\bibitem{Connell_2008}
A.~D. O’Connell {\em et~al.},
\newblock Applied Physics Letters {\bf 92}, 112903 (2008),
  https://doi.org/10.1063/1.2898887.

\bibitem{Tinkham_1994}
M.~Tinkham,
\newblock {\em Introduction to Superconductivity}, {2nd} ed. ({McGraw-Hill},
  New York, 1994).

\bibitem{Leduc_2010}
H.~G. Leduc {\em et~al.},
\newblock Applied Physics Letters {\bf 97}, 102509 (2010),
  https://doi.org/10.1063/1.3480420.

\bibitem{Barends_Lifetime_2008}
R.~Barends {\em et~al.},
\newblock Phys. Rev. Lett. {\bf 100}, 257002 (2008).

\bibitem{Kaplan_1976}
S.~B. Kaplan {\em et~al.},
\newblock Phys. Rev. B {\bf 14}, 4854 (1976).

\bibitem{deVisser2012}
P.~J. de~Visser {\em et~al.},
\newblock Journal of Low Temperature Physics {\bf 167}, 335 (2012).

\bibitem{pozar2011microwave}
D.~Pozar,
\newblock {\em Microwave Engineering, 4th Edition} (Wiley, 2011).

\end{thebibliography}
\end{document}